# Toward Ethical AI: A Qualitative Analysis of Stakeholder Perspectives

Ajay Kumar Shrestha
*Computer Science Department*
*Vancouver Island University*
Nanaimo, Canada
ajay.shrestha@viu.ca

Sandhya Joshi
*VIU Affiliate*
*Vancouver Island University*
Nanaimo, Canada
sandhya.joshi@viu.ca

*Abstract*—As Artificial Intelligence (AI) systems become increasingly integrated into various aspects of daily life, concerns about privacy and ethical accountability are gaining prominence. This study explores stakeholder perspectives on privacy in AI systems, focusing on educators, parents, and AI professionals. Using qualitative analysis of survey responses from 227 participants, the research identifies key privacy risks, including data breaches, ethical misuse, and excessive data collection, alongside perceived benefits such as personalized services, enhanced efficiency, and educational advancements. Stakeholders emphasized the need for transparency, privacy-by-design, user empowerment, and ethical oversight to address privacy concerns effectively. The findings provide actionable insights into balancing the benefits of AI with robust privacy protections, catering to the diverse needs of stakeholders. Recommendations include implementing selective data use, fostering transparency, promoting user autonomy, and integrating ethical principles into AI development. This study contributes to the ongoing discourse on ethical AI, offering guidance for designing privacy-centric systems that align with societal values and build trust among users. By addressing privacy challenges, this research underscores the importance of developing AI technologies that are not only innovative but also ethically sound and responsive to the concerns of all stakeholders.

## I. INTRODUCTION

Artificial Intelligence (AI) systems are rapidly transforming various aspects of society, offering groundbreaking opportunities across domains such as education, healthcare, and technology [1], [2], [3]. From personalized learning paths for students to advanced diagnostic tools in medicine, AI has the potential to improve lives significantly. However, these advancements come with growing concerns about data privacy, security, and ethical accountability, particularly as AI systems increasingly rely on the collection and processing of personal information [4], [5].

The issue of privacy within AI systems is especially pressing in contexts involving sensitive populations, such as children, students, and other vulnerable groups [6]. Parents and educators express apprehensions about how these systems handle data, fearing unauthorized access, ethical misuse, and potential harm [7], [8]. Simultaneously, AI professionals face the technical and ethical challenges of balancing innovation with robust privacy protections, often navigating the fine line between leveraging data for system improvement and safeguarding user trust [9].

Understanding the diverse perspectives of these key stakeholders including educators, parents, and AI professionals, is critical for addressing privacy concerns and fostering ethical AI development. Each group brings unique insights and priorities: parents focus on safeguarding children's data, educators emphasize balancing privacy with educational outcomes, and AI professionals prioritize technical solutions to minimize risks while enabling innovation [9], [10]. However, existing literature often falls short in integrating these distinct perspectives into a comprehensive framework, leaving critical gaps in understanding how privacy challenges in AI systems impact diverse stakeholders.

This study seeks to address these gaps by exploring stakeholder perspectives through qualitative analysis, complementing the quantitative findings detailed in a separate study accepted for publication at the same conference [11]. While quantitative study provided a broad overview of privacy-related constructs and their interrelationships, it lacked the depth needed to capture the nuanced experiences, concerns, and priorities of different stakeholder groups. The qualitative approach enables a richer understanding of the problem domain by uncovering contextual details, exploring underlying concerns, and identifying actionable recommendations that may not emerge from quantitative methods alone.

By focusing on qualitative insights, this study delves deeper into stakeholder perspectives, identifying key risks, perceived benefits, privacy concerns, and proposed measures for enhancing privacy in AI systems. Through this approach, the voices of those directly impacted by and involved in AI system development are integrated, contributing to the creation of technologies that are not only innovative but also ethically sound and privacy-centric. The findings bridge the gap in literature, providing a foundation for designing AI systems that align with societal values, foster trust, and meet the needs of diverse stakeholders in a more holistic manner.

In the next section, we present background and related works, including prior research on privacy concerns in AI systems and existing frameworks. Section III describes the methodology, including the survey design, participant demographics, and the thematic analysis framework employed to analyze qualitative responses. The results, detailing stakeholder perspectives on privacy risks, benefits, concerns, and proposed measures, are presented in Section IV. Section V discusses the implications for stakeholders and ethical AI development. Finally, Section VI concludes the paper with key contributions, limitations, and directions for future research.

## II. BACKGROUND AND RELATED WORKS

The growing integration of Artificial Intelligence (AI) systems into daily life has heightened concerns about privacy and ethical practices. AI systems rely on vast amounts of personal data to deliver personalized experiences and improve decision-making processes. However, this reliance introduces significant risks, especially in sectors like education, where sensitive data about young users is frequently collected and processed.

### A. Privacy Concerns in AI Systems for Young Digital Citizens

Young digital citizens, defined as individuals raised in technology-driven environments face distinct privacy challenges arising from their interactions with AI-powered tools and applications. These challenges include the extensive collection and processing of personal data, which, while enabling tailored experiences, increases risks of data breaches and unauthorized use [12], [13]. Studies have shown that younger users often lack awareness of how their data is utilized, making them more vulnerable to privacy violations and unethical data practices [7], [14]. For example, research involving caregivers and youth highlights that parental mediation strategies significantly influence youth privacy awareness, while a lack of guidance exacerbates risks [15]. Additionally, high-profile data breaches and unethical practices on platforms that leverage AI, such as social media, have amplified concerns among parents and educators [16]. Despite their vulnerability, young digital citizens are not often directly involved in shaping AI privacy strategies, pointing to a gap in understanding and addressing their unique needs.

### B. Key Stakeholder Perspectives

Parents and educators, as mediators of technology use for young digital citizens, play a critical role in ensuring digital safety and fostering awareness of privacy practices. Their trust in AI systems depends heavily on transparent communication about data usage and strong privacy protections [16], [17]. Research shows that clear, accessible explanations about AI data practices can significantly influence their comfort in promoting the use of such systems among young users [7], [18]. AI developers and researchers, on the other hand, are responsible for designing systems that balance functionality, innovation, and ethical considerations. Their work often involves decisions about data collection, processing, and storage that directly impact user privacy. Studies suggest that cultural and social contexts also influence trust in AI systems, adding complexity to how developers address privacy in global applications [19], [20]. While technical stakeholders acknowledge the importance of data control and user consent, their priorities frequently center on system performance and innovation, creating a gap between their focus and the concerns of non-technical stakeholders [21], [22].

### C. Existing Approaches to Address Privacy Concerns

Several approaches have been proposed to mitigate privacy risks in AI systems. Privacy-by-design principles advocate for embedding privacy measures, such as data anonymization, encryption, and minimization, into the architecture of AI systems from the outset [23]. Federated learning, which trains AI models locally on devices rather than centralizing data, is another emerging technique to reduce data exposure [24], [25], [26]. Additionally, regulatory frameworks such as the Personal Information Protection and Electronic Documents Act (PIPEDA) in Canada have set standards for transparency, accountability, and user consent in data processing. Despite these advancements, gaps remain in addressing the unique privacy needs of diverse stakeholders.

### D. Gaps in Existing Research and the Role of Qualitative Analysis

While prior studies have highlighted the technical and regulatory aspects of privacy in AI systems, there is limited research exploring the perspectives of non-technical stakeholders such as parents and educators. Furthermore, existing work often overlooks the practical challenges and trade-offs faced by AI professionals when balancing privacy protection with system functionality. This study aims to fill these gaps by providing a comprehensive analysis of stakeholder perspectives, integrating their views into actionable recommendations for ethical AI development. By situating this study within the broader context of privacy research in AI, the findings contribute to advancing the discourse on how AI systems can align with societal values and user expectations while addressing the complexities of privacy protection. This foundation sets the stage for the methodological approach and insights presented in the subsequent sections.

## III. METHODOLOGY

In this section, we outline our research goal, questions, processes, instrumentation and the coding methodology.

### A. Goal and Research Questions

The primary goal of this study was to investigate stakeholder perspectives on privacy in AI systems, focusing on educators, parents, and AI professionals. The study sought to address the following research questions, which form the foundation of the conceptual framework illustrated in Fig. 1:

- RQ1: What are the primary risks associated with the use of personal data in AI systems?
- RQ2: What benefits do stakeholders perceive from AI systems using personal data?
- RQ3: What are the main privacy concerns regarding AI systems?
- RQ4: What measures should be implemented to enhance privacy in AI systems?
- RQ5: How can the benefits of data usage be balanced with privacy protection?

Fig. 1 provides a visual representation of how these research questions are mapped to the five themes derived from stakeholder feedback: Primary Risks, Perceived Benefits, Privacy Concerns, Proposed Measures, and Balancing Benefits and Privacy. These themes, informed by the research questions, converge to inform actionable insights for Ethical AI Development. The definitions of these themes are provided in Table I, which details their scope and focus within the study.

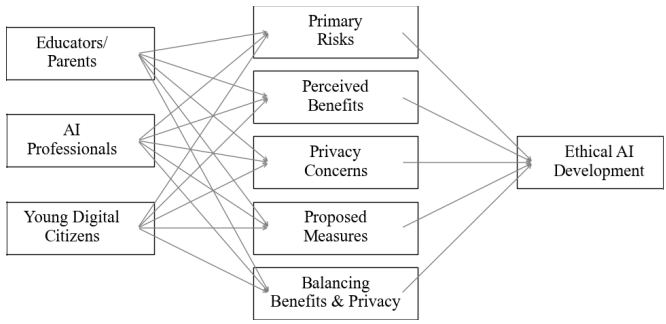

Fig. 1. Stakeholder Perspectives Leading to Ethical AI Development

TABLE I. THEMES AND DEFINITION

| Theme | Definition |
|---|---|
| **Primary Risks** | Potential dangers from personal data use, such as breaches, misuse, and ethical concerns. |
| **Perceived Benefits** | Advantages of AI using personal data, including personalized services and innovation. |
| **Privacy Concerns** | Worries about data security, transparency, and surveillance. |
| **Proposed Measures** | Suggestions to improve privacy, like transparency, anonymization, and encryption. |
| **Balancing benefits and privacy** | Strategies to optimize data use while safeguarding privacy. |

*B. Research Process*

The present study received ethics approval from the Vancouver Island University Research Ethics Board (VIU-REB). The approval with reference number #103116 was given for behavioral application/amendment forms, consent forms and questionnaires. We conducted a pilot study with 6 participants, including members of empirical research specialists from the University of Saskatchewan and Vancouver Island University. The pilot study aimed to assess the feasibility and duration of the research approach and refine the study design. Participants provided general feedback on the questionnaire which informed modifications and restructuring of the final survey questionnaires. The revised research model was then tested by gathering survey data.

We recruited participants through flyers, emails, personal networks, and on social networking sites, LinkedIn and Reddit. Participation was entirely voluntary and did not receive any form of compensation. The participants had to read and accept a consent form to participate in the study, by submitting the consent form before starting the questionnaire participants were indicating they understood the conditions of participation in the study outlined in the consent form. We conducted online surveys through Microsoft Forms by requesting each participant to respond to the questionnaire based on our three designated demographics: AI Researchers and Developers, Parents and Educators, and Young Digital Citizens aged 16-19. Young digital citizens were excluded from this paper due to their small sample size and misalignment with the study's focus on stakeholder perspectives, but they will be included in future analyses.

A total of 252 participants were initially surveyed. After data cleaning to remove incomplete or irrelevant responses, the valid sample included:

*1) Educators/Parents:* 110 valid responses. Within this group:
    *a)* 41 identified as parents.
    *b)* 44 identified as educators.
    *c)* 25 identified as both educators and parents.

*2) AI Professionals:* 117 valid responses. Within this group:
    *a)* 40 identified as AI developers.
    *b)* 76 identified as AI researchers.
    *c)* 1 identified as both a developer and a researcher.

*C. Instrumentation and Coding*

The survey instrument consisted of five open-ended questions designed to elicit qualitative responses regarding privacy in AI systems. Questions were structured to address the research objectives, focusing on risks, benefits, privacy concerns, proposed measures, and the balance between privacy and data usage.

Responses were thematically analyzed using an inductive approach. The coding process included the following steps:

*1) Familiarization:* Researchers thoroughly reviewed all responses to gain a comprehensive understanding of the data.

*2) Initial Coding:* Open coding was conducted to identify recurring concepts and themes.

*3) Theme Development:* Codes were grouped into broader themes that aligned with the study's objectives. The following themes emerged from the analysis:
    *a) Primary Risks:* Privacy breaches, ethical concerns, and misuse of data.
    *b) Perceived Benefits:* Personalized experiences, improved services, educational advancements, and innovation.
    *c) Privacy Concerns:* Excessive data collection, lack of transparency, and profiling.
    *d) Proposed Measures:* Enhanced transparency, data anonymization, privacy-by-design, and ethical oversight.
    *e) Balancing Benefits and Privacy:* Selective data use, user empowerment, and proportionality in data collection.

*4) Validation:* Themes were reviewed by a second researcher to ensure consistency and accuracy.

*5) Finalization:* A consolidated codebook was created, and themes were finalized for integration into the analysis.

## IV. RESULTS

We processed the data collected through Microsoft Forms using Microsoft Excel and performed thematic analysis on the open-ended responses to identify key themes and participant concerns.

*A. Primary Risks Associated with Personal Data in AI Systems (RQ1)*

Participants highlighted several critical risks associated with the use of personal data in AI systems. A prominent concern was the vulnerability of these systems to privacy breaches, with stakeholders emphasizing the potential for unauthorized access and misuse of sensitive information. Parents and educators were particularly alarmed by the risks posed to young people, citing fears of long-term harm resulting from compromised data security. As one parent explained, "Unauthorized access to personal data can harm children in ways they don't even understand yet" (Participant #12). AI professionals echoed these concerns, pointing to systemic vulnerabilities within security frameworks that could leave large-scale datasets exposed. One participant noted, "Systems are not designed to handle the scale of potential breaches" (Participant #45).

Ethical concerns also emerged as a recurrent theme, with participants raising issues related to biased decision-making, lack of informed consent, and the exploitation of personal data. Parents and educators expressed unease about the opaque nature of AI systems and their potential to reinforce societal inequities. As one educator observed, "AI systems are a black box when it comes to ethics and consent" (Participant #78). AI professionals, on the other hand, underscored the broader implications of deploying AI systems without robust ethical safeguards, highlighting risks such as systemic discrimination and the erosion of public trust in technology.

Another significant risk identified by participants was the misuse of personal data, particularly in the context of targeted misinformation, manipulation, and discrimination. Parents voiced fears about the exploitation of their children's data for purposes beyond their consent, such as targeted advertising or political influence. "Data misuse can lead to manipulation and harm, especially for kids," one parent remarked (Participant #23). AI professionals acknowledged the challenges of mitigating such risks, pointing out the complexities of ensuring data security in expansive, interconnected ecosystems. One participant warned, "Once data is shared, there's no going back. Misuse becomes inevitable" (Participant #90).

Table II summarizes these findings by categorizing the identified risks across the three stakeholder groups, parents, educators, and AI professionals, highlighting the commonalities and differences in their concerns. While all stakeholders are aligned on the importance of privacy and security, parents and educators focus more on the implications for children and students, whereas AI professionals are concerned with systemic risks and ethical safeguards. This alignment of concern underscores the urgent need for privacy-centric policies and ethical practices tailored to diverse stakeholder needs.

TABLE II. IDENTIFIED RISKS OF PERSONAL DATA USE IN AI SYSTEMS BY STAKEHOLDER GROUP

|  | Parents | Educators | AI Professionals |
|---|---|---|---|
| **Privacy Breaches** | Unauthorized access to children's data | Insecure systems for student information | Systemic vulnerabilities in large datasets |
| **Ethical Concerns** | Lack of consent for child-specific tools | Black-box decision-making processes | Systemic bias and lack of ethical checks |
| **Misuse of Data** | Exploitation for manipulation or advertising | Overuse of educational data | Unintended consequences of data misuse |

## B. Perceived Benefits of AI Systems Using Personal Data (RQ2)

Participants highlighted several significant benefits of AI systems that leverage personal data. One of the most frequently mentioned advantages was the ability of AI to provide personalized experiences. Parents and educators appreciated AI's capacity to tailor educational tools and resources to meet individual learning needs. As one parent noted, "AI tools can personalize learning for children, helping them grasp concepts more effectively" (Participant #15). AI professionals emphasized that personalization extends beyond education, improving user experiences in healthcare, entertainment, and customer service. One professional stated, "Tailored recommendations are one of the greatest strengths of AI systems" (Participant #22).

Another recurring theme was the improvement of services enabled by AI systems. Many respondents acknowledged the potential of AI to enhance efficiency and functionality across various domains. Educators highlighted how AI-driven tools simplify lesson planning and assessment, while parents recognized the benefits of educational advancements for their children. One educator commented, "With AI, I can easily identify areas where my students struggle and provide targeted help" (Participant #65). AI professionals also pointed out that improved services make essential resources, such as healthcare and financial tools, more accessible to broader populations.

Educational advancement was another benefit discussed, particularly by educators. AI's ability to create adaptive teaching tools and foster engagement among students was highly valued. These tools were viewed as instrumental in making learning more inclusive and effective. As one educator explained, "AI can make learning more engaging and accessible for students with diverse needs" (Participant #35). Parents similarly expressed optimism about AI's role in preparing children for a technology-driven future.

Lastly, participants, particularly AI professionals, highlighted the role of AI systems in driving innovation. They described how personal data fuels the development of smarter algorithms and helps solve complex global challenges. Examples cited included climate modeling, medical research, and resource optimization. One AI professional remarked, "Data-driven innovation is unlocking solutions to some of the world's toughest problems" (Participant #50). Table III summarizes the diverse benefits perceived by stakeholders.

TABLE III. BENEFITS OF AI SYSTEMS BY STAKEHOLDER GROUP

|  | Parents | Educators | AI Professionals |
|---|---|---|---|
| **Personalized Experiences** | Tailored learning for children | Individualized student resources | Customized recommendations |
| **Improved Services** | Enhanced tools for child education | Simplified lesson planning | Streamlined processes across domains |
| **Educational Advancement** | Preparing children for the future | Inclusive and adaptive tools |  |
| **Innovation** |  |  | Solving complex challenges through data-driven insights |

While parents and educators emphasized educational and personalized applications, AI professionals focused on broader societal advancements and innovation. These findings illustrate the multifaceted potential of AI systems when personal data is used responsibly and ethically.

*C. Main Privacy Concerns Regarding AI Systems (RQ3)*

Participants expressed a range of privacy concerns regarding AI systems, which centered around three primary themes. One significant issue was the accuracy of predictions and recommendations generated by AI systems. Stakeholders raised concerns about how reliable these outputs are, particularly when the data lacks context or contains biases. Parents highlighted the potential for harm caused by inaccurate predictions, especially when such outputs influence critical decisions. As one parent noted, "AI predictions often lack context, making them unreliable for decision-making" (Participant #45). AI professionals echoed this concern, emphasizing the importance of building trust in AI systems through reliable and accurate predictions. One participant remarked, "How accurate is the prediction and recommendation? Inaccuracy can harm trust and outcomes" (Participant #72).

Another pressing issue identified by participants was excessive data collection. Many stakeholders raised alarms about the tendency of AI systems to collect more data than necessary, often without clear user consent or understanding. Educators expressed frustration about the overcollection of data in educational settings, which they perceived as unnecessary for fulfilling the intended purpose of AI tools. One educator stated, "AI systems ask for too much data, more than they need to serve their purpose" (Participant #31). Parents also voiced concerns about children's inability to fully grasp the implications of sharing personal information, which makes them vulnerable to overcollection. As one parent explained, "Kids don't understand or manage their privacy well, which leads to overcollection of data" (Participant #27).

The third theme revolved around user profiling and surveillance, with participants highlighting the risks associated with data aggregation and the creation of detailed user profiles. Stakeholders feared that such profiling could lead to misuse, including targeted advertising, manipulation, and unwarranted surveillance. Educators expressed concern about how collected data could be repurposed for objectives beyond its original intent, potentially infringing on privacy. One educator remarked, "Data captured by AI systems can be manipulated to serve purposes beyond their original intent" (Participant #18). AI professionals also warned about the ethical and societal risks posed by profiling, including discrimination and misinformation. One participant stated, "Profiling creates risks of discrimination and misinformation" (Participant #65).

Table IV highlights the shared concerns across stakeholder groups, with parents and educators focusing on protecting children and students, and AI professionals emphasizing the technical and societal implications of inaccurate predictions and profiling. Addressing these concerns requires a combination of technical safeguards, ethical oversight, and transparent practices to ensure user trust and data integrity.

TABLE IV. MAIN PRIVACY CONCERNS IN AI SYSTEMS BY STAKEHOLDER GROUP

|  | Parents | Educators | AI Professionals |
|---|---|---|---|
| **Accuracy of Predictions and Recommendations** | Unreliable predictions affecting decisions | Biases in outputs affecting learning outcomes | Concerns about targeting and manipulation |
| **Excessive Data Collection** | Vulnerability of children to overcollection | Collection of unnecessary student data | Fear of repurposing data beyond the original intent |
| **User Profiling and Surveillance** | Concerns about targeting and manipulation | Risks of overcollection without consent | Ethical risks of discrimination and misinformation |

*D. Measures to Enhance Privacy in AI Systems (RQ4)*

Participants suggested several actionable measures to address privacy concerns in AI systems, highlighting the need for enhanced transparency, technical safeguards, user empowerment and the potential of emerging technologies like smart contracts. A recurring theme was the importance of enhanced transparency, where stakeholders emphasized the necessity of clear, accessible, and detailed communication about how data is collected, stored, and used. Parents particularly valued transparency for building trust, with one parent stating, "Transparency about what data is collected and why would build trust with parents" (Participant #22). AI professionals echoed this, noting that providing comprehensive data usage policies could alleviate many concerns (Participant #56). Smart contracts were suggested as a potential tool to operationalize transparency by automating and documenting data-sharing agreements, ensuring that all parties have a clear understanding of how data will be used. Participant #128 further highlighted their potential, stating, "Maybe even use differential privacy or smart contracts on blockchain to give users control over their own data."

Another key recommendation was the implementation of data anonymization to mitigate the risks of misuse and unauthorized identification. Educators and AI professionals alike advocated for anonymizing personal data to safeguard user identities while enabling useful applications. As one educator noted, "If data must be collected, it should be anonymized to protect identities" (Participant #40). Similarly, an AI professional added, "Anonymization ensures that even if data is accessed, it cannot harm individuals" (Participant #73).

Participants also proposed local AI training to enhance privacy. Training AI models on local devices, rather than central servers, was viewed to reduce exposure to potential breaches. This approach aligns with federated learning principles, which limit data sharing while maintaining system functionality. One educator remarked, "Train AI models on local devices using a federated learning approach to limit data sharing" (Participant #88), a sentiment supported by an AI professional who emphasized its potential to reduce security vulnerabilities (Participant #101). Smart contracts were noted by one researcher (Participant #127) as a complementary tool for managing distributed data-sharing agreements securely, further enhancing privacy protection. The researcher explained,

"Smart contracts could even help enforce privacy rules automatically," highlighting their role in reinforcing trust and accountability.

The use of encryption and deletion was another frequently mentioned measure. Stakeholders emphasized encrypting sensitive information to protect it during storage and transit, as well as ensuring the deletion of raw data after analysis to prevent future misuse. A parent noted, "Encrypt sensitive information and delete raw data to minimize risks" (Participant #54), while an AI professional advocated for end-to-end encryption as a standard practice (Participant #76).

Finally, participants stressed the need for user empowerment by providing tools that allow users to control, monitor, and adjust their privacy settings effectively. Parents highlighted the importance of having decision-making tools, with one stating, "Parents should have tools to decide how their children's data is used" (Participant #18). Educators also saw user empowerment as a way to build accountability in data sharing, as noted by one participant: "Giving users more control over data sharing builds accountability" (Participant #31). Table V summarizes the diverse priorities of different stakeholder groups.

### E. Balancing Benefits and Privacy Protection (RQ5)

Participants underscored the critical need to balance the benefits of data usage in AI systems with robust privacy protection measures. A common theme was the importance of selective data use, with stakeholders advocating minimizing the amount of data collected to what is strictly necessary for achieving system objectives. Parents emphasized the importance of limiting data sharing for essential purposes only. As one parent explained, "We only allow the minimum amount of data to be shared for essential purposes" (Participant #29). AI professionals echoed this sentiment, emphasizing the value of proportional data collection in maintaining system performance and privacy. One participant remarked, "Collecting only relevant data ensures privacy without compromising system performance" (Participant #40).

Another essential measure identified by participants was the need for clear and transparent communication about data usage policies, objectives, and their specific purposes.

Stakeholders viewed clear, open communication as foundational for building trust and enabling informed decision-making. Educators highlighted the importance of explaining data use to parents and other stakeholders, particularly in educational settings. One educator stated, "We explain to parents how student data will be used to improve learning outcomes" (Participant #72). Similarly, AI professionals recognized transparency to foster trust among users. As one professional noted, "Transparency in how we handle data fosters trust among stakeholders" (Participant #93).

The principle of privacy-by-design emerged as a proactive approach to ensuring data security while enabling meaningful use of information. Stakeholders agreed that privacy protections should be embedded into AI systems from the outset rather than being addressed retrospectively. A parent articulated this by stating, "Systems should be built with privacy protections from the start, not as an afterthought" (Participant #12). AI professionals also emphasized the importance of designing systems that integrate privacy safeguards, with one stating, "Privacy by design ensures data security while enabling meaningful use of information" (Participant #65).

Finally, participants stressed the necessity of ethical oversight in achieving a fair balance between data usage and privacy. They advocated for the establishment of ethical guidelines and the involvement of independent oversight bodies to ensure responsible data practices. Educators highlighted the role of audits and policies in preventing data exploitation. One educator noted, "Policies and audits help ensure data is used responsibly without exploiting individuals" (Participant #88). AI professionals similarly emphasized the importance of aligning data practices with societal values through ethical oversight. One professional remarked, "Ethical oversight prevents misuse and aligns data practices with societal values" (Participant #101).

Table VI highlights the shared commitment across stakeholder groups to balancing the utility of data with strong privacy protections. Parents and educators emphasized safeguarding children's and students' data, while AI professionals focused on technical and ethical mechanisms to uphold privacy while enabling innovation. Together, these strategies offer a roadmap for designing AI systems that balance benefits and privacy responsibly.

TABLE V. MEASURES TO ENHANCE PRIVACY IN AI SYSTEMS BY STAKEHOLDER GROUP

|  | Parents | Educators | AI Professionals |
|---|---|---|---|
| Enhanced Transparency | Clear communication about data collection | Open communication about data use | Detailed data usage policies for users |
| Data Anonymization | Protect children's identities | Anonymize student information | Privacy through anonymization |
| Local AI Training | Risks of large-scale data accumulation | Train models locally to protect data | Federated learning to enhance security |
| Encryption and Deletion | Encrypt and delete sensitive data |  | End-to-end encryption standards |
| User Empowerment | Tools to manage children's data | Tools for privacy management | User-controlled data sharing; supported by smart contracts |

TABLE VI. STRATEGIES FOR BALANCING BENEFITS AND PRIVACY PROTECTION BY STAKEHOLDER GROUP

|  | Parents | Educators | AI Professionals |
|---|---|---|---|
| Selective Data Use | Limit data sharing to essential purposes | Collect minimal data for student performance | Collect only relevant data |
| Transparent Communication | Clear explanation of data-sharing practices | Explain educational benefits to parents | Transparent policies for all stakeholders |
| Privacy-By-Design | Build systems with embedded protections | Privacy-integrated learning tools | Integrate privacy safeguards from the start |
| Ethical Oversight | Policies ensuring no misuse of children's data | Audits and ethical policies for accountability | Independent review boards to guide decisions |

## V. Discussion

The findings from this study highlight critical considerations for the development of ethical AI systems, particularly in the context of privacy protection and stakeholder engagement. Grounded in participants' responses, these implications provide actionable insights for fostering trust, accountability, and inclusiveness in AI development. This qualitative analysis complements the quantitative findings detailed in a separate study [11]. The key implications are discussed below:

### A. Prioritizing Transparency and Communication

Transparency emerged as a cornerstone for building trust among stakeholders. Developers and organizations must prioritize creating clear and accessible communication channels about data collection, storage, and usage practices. Transparency involves not only sharing policies but also ensuring they are understandable to diverse stakeholder groups, including non-technical audiences such as parents and educators. For instance, one parent noted, "Transparency about what data is collected and why would build trust with parents" (Participant #22). Involving educators in co-creating policies for educational AI tools could foster broader acceptance and trust.

### B. Embedding Privacy-By-Design Principles

The concept of privacy-by-design calls for integrating privacy considerations into AI systems from the outset. This approach ensures that privacy protection is not an afterthought but a fundamental component of the system's architecture. Developers should incorporate features like data minimization, encryption, anonymization, and local AI training to reduce vulnerabilities and enhance user trust. As one educator suggested, "If data must be collected, it should be anonymized to protect identities" (Participant #40). Privacy-by-design principles can serve as a competitive differentiator, helping organizations align with evolving regulatory frameworks and societal expectations.

### C. Strengthening Ethical Oversight

Ethical oversight mechanisms, such as independent review boards, can help ensure AI systems adhere to ethical principles, including fairness, accountability, and transparency. Regular audits of data usage practices and model performance can mitigate biases, prevent exploitation, and foster equitable outcomes. Participants frequently highlighted the need for such mechanisms, with one AI professional stating, "Ethical oversight prevents misuse and aligns data practices with societal values" (Participant #101). Ensuring that educators and parents have representation on review boards for educational AI systems can promote inclusivity and accountability.

### D. Promoting User Empowerment

Participants stressed the need for AI systems to empower users with tools to control their data and make informed decisions. Features such as adjustable privacy settings, detailed consent mechanisms, and real-time data usage dashboards can enhance user autonomy. For parents and educators, these tools are particularly important to safeguard the interests of children and young learners. One parent remarked, "Parents should have tools to decide how their children's data is used" (Participant #18). Empowering users not only addresses privacy concerns but also encourages active engagement and participation in AI-driven ecosystems.

### E. Balancing Innovation and Privacy

While leveraging the transformative benefits of AI systems, it is essential to strike a delicate and thoughtful balance between fostering innovation and upholding privacy. Developers must rigorously evaluate the necessity and proportionality of data collection, ensuring that data-driven advancements are pursued without eroding user trust or compromising their safety. This involves not only minimizing data collection to what is strictly relevant but also embedding privacy considerations into every stage of the AI development lifecycle. As one AI professional noted, "Collecting only relevant data ensures privacy without compromising system performance" (Participant #40). Ethical AI development demands that organizations align their business objectives with societal values, prioritizing transparency, accountability, and the equitable treatment of all stakeholders. By fostering a culture of responsibility and ethical leadership, developers and organizations can ensure that AI innovations contribute positively to society while maintaining the highest standards of privacy and trust.

## VI. Conclusion

This study offers a detailed examination of stakeholder perspectives on privacy in AI systems, with a focus on educators, parents, and AI professionals. Through thematic analysis of qualitative data, the research identified key risks, including privacy breaches, ethical concerns, and data misuse, as well as notable benefits such as personalized experiences, improved services, and educational advancements. Stakeholders highlighted pressing privacy concerns, ranging from excessive data collection to inadequate transparency, and proposed actionable measures like enhanced transparency, data anonymization, privacy-by-design, and ethical oversight. Balancing these privacy concerns with the benefits of AI emerged as a central challenge, emphasizing the importance of selective data use, user empowerment, and ethical innovation. The study's findings also revealed how diverse stakeholder groups approach privacy issues, with parents prioritizing child safety, educators focusing on learning outcomes, and AI professionals emphasizing data security and innovation. These perspectives underscore the need for AI systems that cater to varied priorities while fostering trust through inclusive design and transparent communication. By addressing these findings, developers and organizations can build ethical AI systems that balance privacy with utility, align with societal values, and encourage broader stakeholder engagement. While the research provides meaningful insights, its limitations, such as the exclusion of young digital citizens, reliance on self-reported data, and lack of longitudinal analysis, must be acknowledged. Future research should address these gaps by incorporating diverse demographics, exploring evolving stakeholder perspectives, and employing mixed methods. This holistic approach can further advance the development of ethical AI systems that prioritize privacy and accountability while leveraging their transformative potential for societal benefit.

ACKNOWLEDGMENT

This project has been funded by the Office of the Privacy Commissioner of Canada (OPC); the views expressed herein are those of the authors and do not necessarily reflect those of the OPC.